# Comparison of alternative zebra-structure models in solar radio emission

Chernov G.P.[1], Fomichev V.V.[1], Sych R.A.[2] ✴

Discussion about the nature of zebra-structure (ZS) in the type IV radio bursts continues, despite the ten proposed models. First of all, this is due to the wide variety of stripes in each new phenomenon, when the explanation of all the fine details by any one mechanism becomes impossible. The most widespread explanation is the emission at different levels of double plasma resonance (DPR), sequential on the height surfaces in the magnetic trap, where the upper hybrid frequency ($\omega_{UH}$) becomes equal to the integer of electronic cyclotron harmonics $s\omega_{Be}: \omega_{UH} = (\omega_{Pe}^2 + \omega_{Be}^2)^{1/2} = s\omega_{Be}$ (Zheleznyakov, Zlotnik, 1975; Winglee, Dulk, 1986; Kuznetsov & Tsap (2007) ). An important alternative mechanism is the interaction of plasma waves with the whistlers: $l + w \Rightarrow t$ (Chernov, 1976; 2006). Here, we will show the possibility of explaining the main features of the zebra stripes in the model with whistlers, using the example of the phenomenon on August 1, 2010.

## Introduction

A number of authors continues to improve mechanism based on the DPR– model. Karlicky and Yasnov (2015), showed that the ZS can be excited within the framework of the DPR in the transition layer only in a narrow height interval with the high density gradient (model Selhost et al. 2008, A&A, 488,1079). Under such conditions it becomes clear, that the DPR- mechanism cannot provide a large number of zebra stripes. In recent work of Benáček, Karlicky, Yasnov (2017), new calculations of the increments of the upper hybrid waves under DPR- conditions with a ring distribution of high-speed electrons with relativistic corrections for different temperatures of background plasma and fast particles and for optimal wave numbers were carried out. It was shown that the explisit maximum of the increment is obtained only for electron velocities = 0.1$c$ with a narrow dispersion (for the speed ~ 0.2 $c$ the increment sharply decreases and the maxima are washed out in the continuum for several cyclotron harmonics $s$). Thus these calculations shows the inefficiency of the DPR– mechanism. It is also important to note that the zebra stripes in this case still reveal the superfine structure of millisecond duration.

**The August 1, 2010 event** is an obvious case to understand all the difficulties in explaining the complex structure of the zebra. Figure 1 gives dynamic spectrum in the decimeter range of Yunan observatory during the first maximum of this phenomenon.

The event on August 1, 2010 was discussed in connecntion with two complex CME. Separate spectra of Ondřejov observatory were used by Marian Karlicky and coauthors in several works under the DPR- model, but no one tried to adequately explain the diversity of ZS in this case (see Figure 1).

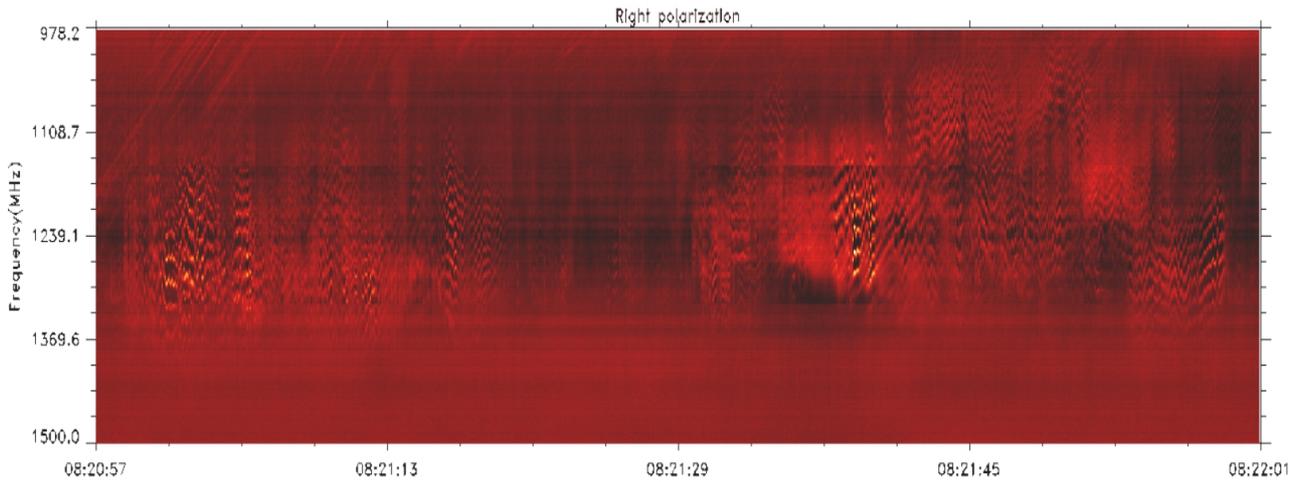

Figure 1. Zebra- structure in the pulsating regime in the event on August 1, 2010 in the decimeter range 1000 – 1500 MHz observed by the spectropolarimeter of Yunan observatory (YNAO, Kunming, China). The radio emission is weakly polarized. (Courtesy Guannan Gao).

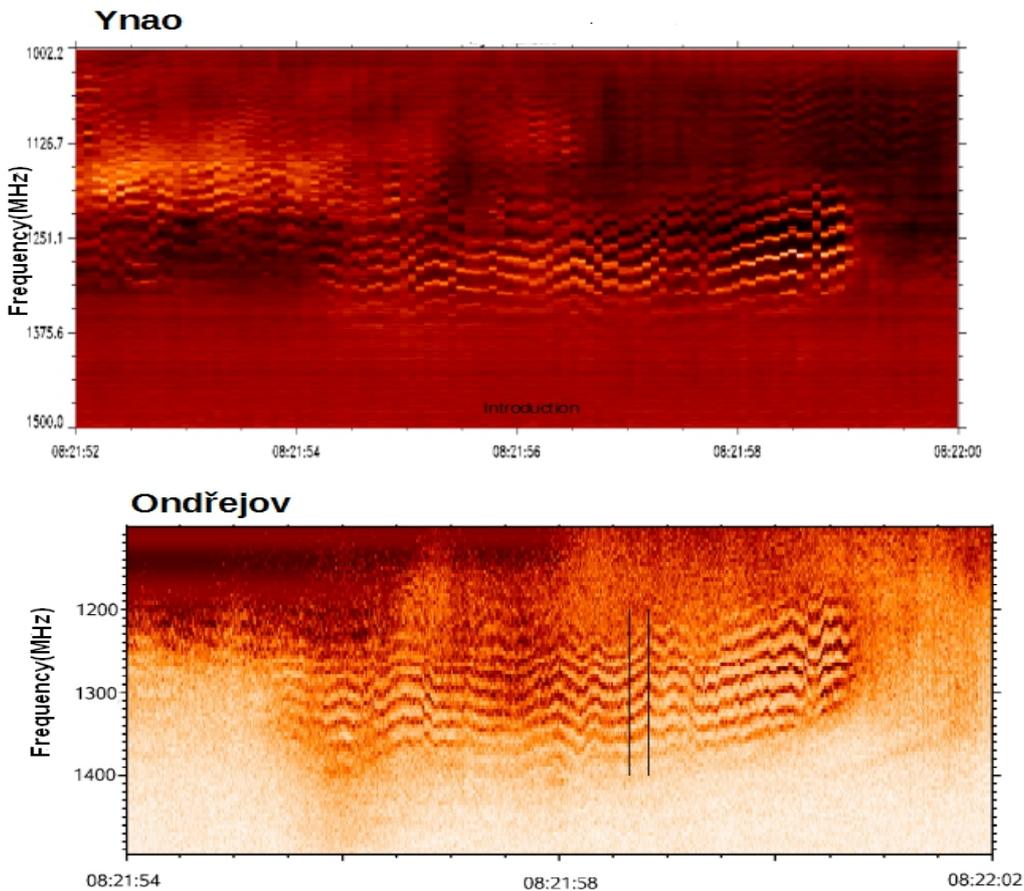

Figure 2. An enlarged fragment of the spectrum shown in Figure 1, of 8 s duration, compared with the spectrum of Ondřejov observatory. All the components of the stripes coincide in two observatories, divided into 8000 km that confirms the solar origin of ZS (the time scale of obs. Ynao is late on 1.6 s). The time resolution of spectrum of Ondřejov is by an order higher, and we see therein the superfine structure of the stripes of millisecond duration (Ondřejov spectrum is courtesy Marian Karlicky).

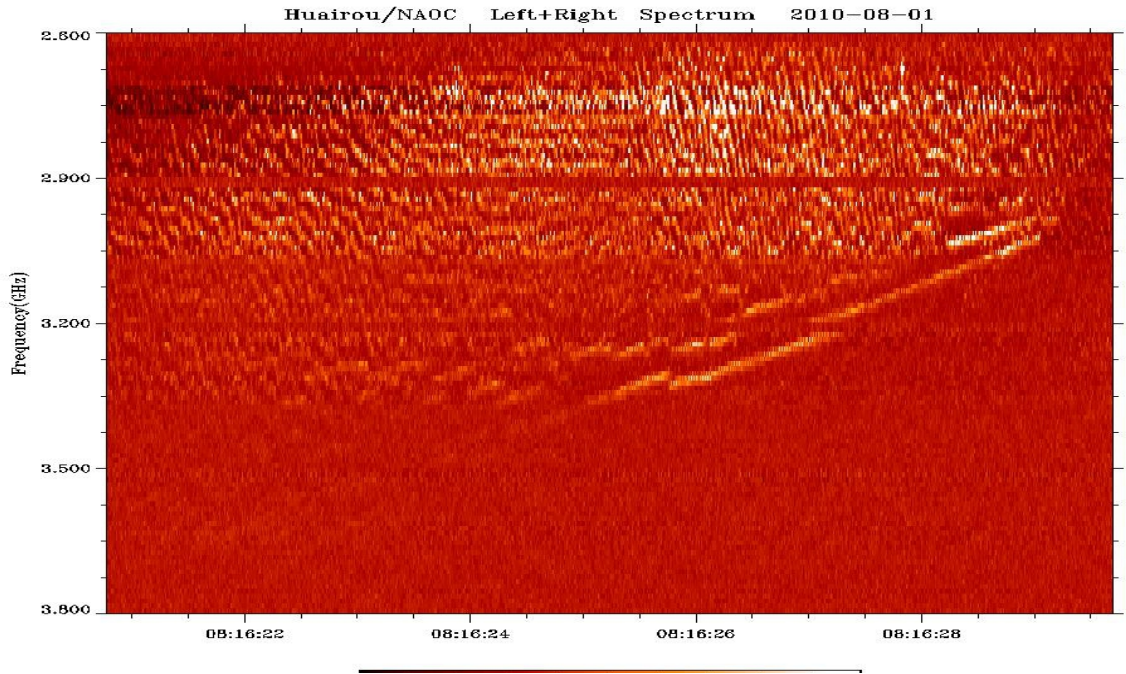

Figure 3. Zebra- structure, limited by fiber bursts from HF edge in the microwave range 2600–3800 MHz of Huairou observatory, Beijing.

**Brief discussion and conclusion**

So, any model should explain:
- radio fibers are sometimes imposed on ZS, or limited it by the LF edge in the decimeter range (or from HF edge in microwave);
- ZS appears in the pulsing mode, chaotic instantaneous columns (almost without their drift) with random duration from 0.1 s to 6 s;
- a wavelike or saw-tooth frequency drift of stripes of ZS in columns;
- the frequency separation grows with the frequency (as usual);
- superfine structure of stripes of ZS and continuum.

From other event we had many examples of frequency splitting of ZS- stripes and, perhaps this was the main property, there was a synchronous change in the frequency drift of ZS stripes with space drift of their radio source.

In the DPR model all the changes of ZS stripes are usually associated with changes of the magnetic field and with propagating fast magneto-acoustic waves. In a very complicated case the simultaneous presence of fast particles in a radio source with several different distribution functions was proposed (Zheleznyakov et al. 2016).

At the same time, in the whistler model all the aforementioned properties of ZS stripes mentioned above were explained by real physical processes occuring during the coalescence of Langmuir waves with whistler waves (Chernov, 2006).

First of all, a radio source of ZS is usually located in magnetic islands after CME ejections. Therefore the close connection of ZS with fiber bursts is simply explained by the acceleration of fast particles in magnetic reconnection regions in the lower part (as in Figure 3) and in the upper part (as in Figure 1) of magnetic islands.

A wavelike or saw-tooth frequency drift of stripes was explained by the switching of whistler instability from the normal Doppler cyclotron resonance into the anomalous one (Fig 2b in Chernov, 1990). Such switching should lead to a synchronous change of the frequency drift of stripes and spatial drift of the radio source, since whistlers generated at normal and anomalous resonance move in opposite directions. New injections of fast particles cause sharp change in frequency drift of stripes in instantaneous ZS columns. Low frequency absorptions (black stripes of ZS) are explained by weakenning of plasma wave instability due to the diffusion of fast particles by whistlers.

The superfine structure is generated by pulsating regime of whistler instability with ion-sound waves (Chernov et al. 2003). Rope-like chains of fiber bursts are explained by periodic whistler instability between two fast shock fronts in a magnetic reconnection region (Chernov, 2006). In the whistler model zebra-stripes can be converted into fiber bursts and back as it is shown in Figure 1.

For a comprehesive discussion of comparative analysis of observations of ZS and fiber bursts and different theoretical models we refer the reader to the reviews of Chernov (2012; 2016) (freely available at http://www.izmiran.ru/~gchernov/).

Acknowledgements

The authors are grateful to the Chinese colleagues in the National Astronomical Observatories of China Chenming Tan and Guannan Gao. The work was supported by the Russian Foundation for Basic Research (Grant N 17-02-00308).

Oral report at the conference: XII Solar System Plasma Conference, February 6–10, 2017, Space Research Institute of RAS.
The full paper was submitted to Astronomy Letters Journal.

∗ [1]Pushkov Institute of Terrestrial Magnetism, Ionosphere and Radio Wave Propagation, Russian Academy of Sciences (IZMIRAN), Troitsk, Moscow 108840, Russia e-mail:gchernov@izmiran.ru
[2]Institute of Solar-Terrestrial Physics of Siberian Branch of Russian Academy of Sciences, 126a Lermontov Street, 664033 Irkutsk, Russia

**References**

Benáček, J., Karlicky, M., Yasnov, L.V. 2017, A&A, 598, A108.
Chernov, G.P. 1990, Sol. Phys. 130, 75.
Chernov, G.P., Yan, Y., Fu, Q. 2003, A&A, 406, 1071.
Chernov, G.P. 2006, Sp. Sci. Rev. 127, 125.
Chernov, G.P. 2011, *Fine structure of solar radio bursts*, *Springer ASSL* 375, Heidelberg.
Chernov, G.P. 2012, In: Reiner A. (Ed)*: Horizons in World Physics,* V. 278, Nova Science Publisher, New-York, Ch. 1, P. 1-75.
Chernov, G.P. 2016, In: Solar Flares: Investigations and Selected Research Editors: Sarah L. Jones, Nova Science Publishers, 2016, Chapter 5: Latest News on Zebra Patterns in the Solar Radio Emission ( pp. 101-149).
Karlicky, M. Yasnov, L.V. 2015, A&A, 581, A115.
Kuznetsov, A. A. & Tsap, Y. T. 2007, Sol. Phys., 241, 127.
Zheleznyakov, V.V., Zlotnik, E.Ya. 1975, Sol. Phys. 44, 461.
Zheleznyakov, V.V., Zlotnik, E. Ya., Zaitsev, V. V., and Shaposhnikov, V. E. 2016, Physics-Uspekhi, 59, N.10, 997.
Winglee, Dulk, 1986, Ap. J, 307, 808.